\documentclass[12pt, draftcls, onecolumn]{IEEEtran}
\usepackage{amssymb}
\usepackage{amsfonts}
\usepackage{amsmath}
\usepackage{cite}
\usepackage[dvips]{graphicx}
\usepackage{color}
\usepackage{comment}
\usepackage{makecell}

\begin{document}

% ===only use in IEEE format : BEGIN ====
\newtheorem{definition}{\it Definition}%[section]
\newtheorem{theorem}{\bf Theorem}%[section]
\newtheorem{lemma}{\it Lemma}
\newtheorem{corollary}{\it Corollary}
\newtheorem{remark}{\it Remark}
\newtheorem{example}{\it Example}
\newtheorem{case}{\bf Case Study}
\newtheorem{assumption}{\it Assumption}
\newtheorem{property}{\it Property}
\newtheorem{proposition}{\it Proposition}
% ===only use in IEEE format : END ====

\newcommand{\hP}[1]{{\boldsymbol h}_{{#1}{\bullet}}}
\newcommand{\hS}[1]{{\boldsymbol h}_{{\bullet}{#1}}}

\newcommand{\ba}{\boldsymbol{a}}
\newcommand{\baq}{\overline{q}}
\newcommand{\bA}{\boldsymbol{A}}
\newcommand{\bb}{\boldsymbol{b}}
\newcommand{\bB}{\boldsymbol{B}}
\newcommand{\bc}{\boldsymbol{c}}
\newcommand{\bcO}{\boldsymbol{\cal O}}
\newcommand{\bh}{\boldsymbol{h}}
\newcommand{\bH}{\boldsymbol{H}}
\newcommand{\bl}{\boldsymbol{l}}
\newcommand{\bm}{\boldsymbol{m}}
\newcommand{\bn}{\boldsymbol{n}}
\newcommand{\bo}{\boldsymbol{o}}
\newcommand{\bO}{\boldsymbol{O}}
\newcommand{\bp}{\boldsymbol{p}}
\newcommand{\bq}{\boldsymbol{q}}
\newcommand{\bR}{\boldsymbol{R}}
\newcommand{\bs}{\boldsymbol{s}}
\newcommand{\bS}{\boldsymbol{S}}
\newcommand{\bT}{\boldsymbol{T}}
\newcommand{\bw}{\boldsymbol{w}}

\newcommand{\balpha}{\boldsymbol{\alpha}}
\newcommand{\bbeta}{\boldsymbol{\beta}}
\newcommand{\bOmega}{\boldsymbol{\Omega}}
\newcommand{\bTheta}{\boldsymbol{\Theta}}
\newcommand{\bphi}{\boldsymbol{\phi}}
\newcommand{\btheta}{\boldsymbol{\theta}}
\newcommand{\bvarpi}{\boldsymbol{\varpi}}
\newcommand{\bpi}{\boldsymbol{\pi}}
\newcommand{\bpsi}{\boldsymbol{\psi}}
\newcommand{\bxi}{\boldsymbol{\xi}}
\newcommand{\bx}{\boldsymbol{x}}
\newcommand{\by}{\boldsymbol{y}}

\newcommand{\cA}{{\cal A}}
\newcommand{\bcA}{\boldsymbol{\cal A}}
\newcommand{\cB}{{\cal B}}
\newcommand{\cE}{{\cal E}}
\newcommand{\cG}{{\cal G}}
\newcommand{\cH}{{\cal H}}
\newcommand{\bcH}{\boldsymbol {\cal H}}
\newcommand{\cK}{{\cal K}}
\newcommand{\cO}{{\cal O}}
\newcommand{\cR}{{\cal R}}
\newcommand{\cS}{{\cal S}}
\newcommand{\dcS}{\ddot{{\cal S}}}
\newcommand{\ds}{\ddot{{s}}}
\newcommand{\cT}{{\cal T}}
\newcommand{\cU}{{\cal U}}
\newcommand{\wt}[1]{\widetilde{#1}}

\newcommand{\mA}{\mathbb{A}}
\newcommand{\mE}{\mathbb{E}}
\newcommand{\mG}{\mathbb{G}}
\newcommand{\mR}{\mathbb{R}}
\newcommand{\mS}{\mathbb{S}}
\newcommand{\mU}{\mathbb{U}}
\newcommand{\mV}{\mathbb{V}}
\newcommand{\mW}{\mathbb{W}}

\newcommand{\uq}{\underline{q}}
\newcommand{\ubq}{\underline{\boldsymbol q}}

\newcommand{\red}[1]{\textcolor[rgb]{1,0,0}{#1}}
\newcommand{\gre}[1]{\textcolor[rgb]{0,1,0}{#1}}
\newcommand{\blu}[1]{\textcolor[rgb]{0,0,1}{#1}}

% paper title
% can use linebreaks \\ within to get better formatting as desired
\title{Multi-operator Network Sharing for Massive IoT} % over Licensed and Unlicensed Bands}

\author{Yong~Xiao, Marwan~Krunz, and Tao~Shu

\thanks{Y. Xiao is with the School of Electronic Information
and Communications at the Huazhong University of Science and Technology, Wuhan, China 430074 (e-mail: yongxiao@hust.edu.cn).

%M. Krunz is with the Department of Electrical and Computer Engineering at the University of Arizona, Tucson, AZ 85710 (e-mail: krunz@email.arizona.edu).

M. Krunz is with the Department of Electrical and Computer Engineering, the University of Arizona, Tucson, AZ 85721 USA, and also with the School of Electrical and Data Engineering, University of Technology Sydney, Ultimo, NSW 2007, Australia (e-mail: krunz@email.arizona.edu).
%and M. Krunz are with the Department of Electrical and Computer Engineering at the University of Arizona, Tucson, AZ (e-mails: xyong.2012@gmail.com and krunz@email.arizona.edu).

T. Shu is with the Department of Computer Science and Software Engineering at the Auburn University, Auburn, AL 36849 (e-mail: tshu@auburn.edu).
}}
% and Zixiang Xiong} %Luiz A. DaSilva}

\maketitle

\begin{abstract}
%The Internet-of-Things (IoT) is the key technology to fulfill 5G's vision of ubiquitous connectivity.
%
A recent study predicts that by 2020, up to 50 billion Internet-of-Things (IoT) devices will be connected to the Internet, straining the capacity of the wireless infrastructure, which has already been overloaded with data-hungry mobile applications. %, such as high-definition video streaming and virtual reality(VR)/augmented reality(AR). 
How to accommodate the demand for both massive-scale IoT devices and high-speed cellular services in the physically limited spectrum without significantly increasing the operational and infrastructure costs is one of the main challenges for operators. {In this article, we introduce a new multi-operator network sharing framework that supports the coexistence of IoT and high-speed cellular services. Our framework is based on the radio access network (RAN) sharing architecture recently introduced by 3GPP as a promising solution for operators to improve their resource utilization and reduce system roll-out cost.} {We evaluate the performance of our proposed framework using real base station location data in the city of Dublin collected from two major operators in Ireland.} Numerical results show that our proposed framework can almost double the total number of supported IoT devices %that can be supported 
and simultaneously coexist with other cellular services. % compared with the case without network sharing. % by sharing their network infrastructure and spectrum resources. 
%% We first review recent 3GPP standards on IoT solutions and introduce the challenges of deploying massive scale IoT in cellular spectrum. %discuss the challenges for massive deployment of the IoT. %We then introduce the
%We describe the main difference between the existing intra-operator spectrum sharing and the LSS, and then compare these technologies with the ULL. We highlight the spectrum sharing strategies and control policies proposed for both LSS and ULL as well as the recent standardization progress and other implementation issues.
%We then
%We then introduce our proposed multi-operator network sharing framework and discuss various design issues when both massive IoT and high-speed cellular services can be supported by operators. 
 % infrastructure deployed by two major telecommunication operators in a major city in Europe. 

\end{abstract}

% Note that keywords are not normally used for peerreview papers.
%\begin{IEEEkeywords}
%Energy harvesting, energy trading, communication networks, game theory, stochastic game, wireless energy transfer.
%\end{IEEEkeywords}

\section{Introduction}
\label{Section_Introduction}

%As the demand for wireless data service continues to grow at an exponential pace, %Cellular operators have consequently requested more spectrum to be allocated to the commercial broadband service.
% In addition, according to the requirement of the Next Generation Mobile Networks Alliance, the 5th generation mobile technology (5G) should support over 1000 times of network throughput supported by 4G.   One promising solution is to allow each operator to be able to access other spectrum that is licensed to other operators or that is unlicensed.

%Internet-of-things (IoT) applications attracted significant interests.    it is expected that mobile

The Internet-of-Things (IoT) is a holistic framework for supporting the communication of intelligent devices and services that are employed in diverse verticals, including e-health, environment control, smart city, and autonomous vehicles. It is considered as the key technology to fulfill 5G's vision of ubiquitous connectivity.  %is a holistic framework for supporting the communications of intelligent devices that are employed in diverse domains (a.k.a. verticals), such as mobile healthcare, manufacturing, transportation, agriculture, smart spaces, and others.
The fast proliferation of IoT applications has been driven by continuous decrease in cost, size, and power consumption of IoT devices and rapidly growing demand for intelligent services.  %, and the fast growing demands on services such as sensors/actuators, autonomous vehicle, and smart devices, have
%has been the driven force of . %It has been predicted that
According to Cisco, by 2020, up to 50 billion IoT devices will be connected to the Internet via cellular networks, generating over \$1.9 trillion in revenue across a wide variety of industries % including  retail, healthcare, manufacturing, transportation,  agriculture, and others
\cite{Cisco2017NetworkIndex}.

%an unprecedented wireless traffic, straining the capacity of limited network resources.

%% Telecommunication operators are responsible for providing wireless solutions on a global scale and therefore are in an excellent position to push for the maturity and large-scale deployment of IoT solutions. %To promote innovation and accelerate adoption, 3GPP has standardized multiple narrowband IoT solutions in cellular systems.
%Major operators throughout the world, including AT\&T, Deutsche Telekom, Orange, and China Mobile, are heavily investing network infrastructures to meet the future demand for IoT services. %Unfortunately, the %
{Because no frequency bands are exclusively allocated to IoT services, IoT devices must share spectrum with other technologies. % including 
%It is expected that IoT will generate an unprecedented wireless traffic, straining the capacity of cellular networks %the already overcrowded networks resource including both the spectrum and infrastructure resources of each operator that has already been overloaded with 
3GPP recently introduces multiple solutions that enable the coexistence of IoT services and regular cellular services.} %It is expected that the massive scale of IoT devices will generate an unprecedented wireless traffic, straining the already overcrowded networks resource including both the spectrum and infrastructure resources of each operator that has already been overloaded with data-hungry applications such as high-definition video streaming and virtual reality(VR)/augmented reality(AR). 
The main challenge for operators is therefore to accommodate the traffic generated by both IoT and fast-growing high-speed cellular services (e.g., enhanced Mobile Broadband (eMBB)) without significantly increasing their operational and infrastructure costs. %To address this challenge, r
Recent 3GPP LTE standards promote the idea of {\em network sharing}, i.e., allowing operators to share radio access network (RAN) resources, including network infrastructure and spectrum, to improve the utilization of individual operator's resources and reduce the system roll-out cost/delay. Recent studies reported that network sharing has the potential to save more than 50\% of the infrastructure cost in 5G deployment for a typical European cellular operator\cite{Samdanis2016NetSlicing}.
Despite its great potential, it is known that network sharing between multiple operators could significantly increase the implementation complexity of wireless systems. In addition, 3GPP's network sharing architecture is mainly introduced to support high-speed data service in which a single operator can temporally access a much wider frequency band to support the high-throughput service requested by a single user equipment (UE). However, IoT devices typically generate low-throughput traffic and their data transmission can be intermittent. How to quickly establish a large number of data connections and allocate the required frequency bands for a massive-scale IoT devices that can be associated with multiple operators is still an open problem. % In particular, most IoT devices will only register with a single operator. How to transfer the IoT traffic between different operators and intermittent

In this article, we propose a novel network sharing framework that allows coexistence of IoT and high-speed data services  across  multiple operators. Our proposed framework is based on the active RAN sharing architecture recently introduced in 3GPP Releases 13-15. We present multiple new design solutions that aim at reducing the implementation complexity of network sharing for IoT applications. Furthermore, we simulate a multi-operator cellular system using actual BS location information obtained from two major telecommunication operators in Ireland. Such trace-driven simulations are used to evaluate the performance of our proposed framework under various practical scenarios. %Our architecture  We first 
The rest of this article is organized as follows. We provide an overview of recent 3GPP solutions on IoT and discuss the challenges for a massive deployment of IoT services in cellular networks. %We then introduce the
%We describe the main difference between the existing intra-operator spectrum sharing and the LSS, and then compare these technologies with the ULL. We highlight the spectrum sharing strategies and control policies proposed for both LSS and ULL as well as the recent standardization progress and other implementation issues.
%We then introduce a novel 
%Our proposed framework is based on the active RAN sharing architecture recently introduced in 3GPP Release 13. 
We then introduce our proposed framework and discuss various design issues. Finally, we present numerical results to demonstrate the potential of our proposed framework.
%We also highlight some of the future research directions at the end of this article.
%To the best of our knowledge, this is the first work that discusses the deployment of IoT solutions on the multi-operator network sharing architecture.

%Extending the operation of LTE into unlicensed spectrum will increase the spectrum resources that are accessible for the cellular operators and therefore has the potential to significantly improve the capacity and throughput of the existing cellular networks. % and alleviate the pressure for the exponential increase of the traffic demand.

%The rest of this article is organized as follows...We will consider
%\begin{figure}
%\centering
%\includegraphics[width=4.9 in]{DynamicEnergyTrading2.pdf}
%\caption{Dynamic energy trading in wireless powered communication networks: we illustrate three possible scenarios of energy trading: 1) energy trading between a power grid and a cellular base station, 2) energy trading between two wireless powered cellular mobile devices, and 3) energy trading between two solar-powered electric vehicles. }
%\label{Figure_DETillustration}
%\end{figure}

\section{Current Solutions and Challenges for IoT}
\begin{comment}
IoT applications can be roughly divided into two types according to different  performance and design requirements: mission-critical and massive-type IoT. The first type emphasizes on high reliability, availability and low latency. Examples include fire/gas alarm, health monitoring, and traffic safety. Massive-type IoT devices focus more on the low-cost, long battery life, and good connectivity which can include smart metering, tracking, agriculture, fleet management. Different types of applications require different spectrum access techniques. %Generally speaking, due to the high demands on the QoS guarantee, mission-critical IoT devices are more appropriate to operate in the licensed bands.

%IoT applications can be operated in both licensed and unlicensed bands.
\end{comment}

\subsection{IoT Solutions of 3GPP}

\begin{table*}[tbp]
\centering
\caption{IoT Solutions in 3GPP Release 13\cite{Shirvanimoghaddam2017NOMAIoT}}
\vspace{-0.1in}
\label{Tabel_IoTSolutions}
%\scriptsize
%\tiny
%\resizebox{\columnwidth}{!}{
\begin{tabular}{lllll}
\hline
%802.11ac
 & EC-GSM-IoT & NB-IoT & eMTC \\
\hline
Frequency & \makecell[l]{850-900 MHz and 1800-\\1900 MHz GSM bands} & \makecell[l]{2G/3G/4G spectrum between 450\\MHz and 3.5 GHz; Sub-2 GHz bands\\are preferred for applications requiring\\ good coverage} & \makecell[l]{Legacy LTE between \\450 MHz and 3.5 GHz} \\
\hline
Bandwidth & 200 kHz & 180 kHz & 1.08 MHz \\
\hline
Maximum Transmit Power & 33 dBm, 23 dBm & 23 dBm, 20 dBm & 23 dBm, 20 dBm  \\
%$3$ & $3$ & $15$ & $63$ & $8$ or $10$ msec   \\
%$4$ & $7$ & $15$ & $1023$ & $2$ or $10$ msec  \\
\hline
\end{tabular}
%\vspace{-0.3in}
\end{table*}

Three solutions have been standardized by 3GPP for cellular IoT deployment: extended coverage GSM IoT (EC-GSM-IoT), narrowband IoT (NB-IoT), and enhanced machine-type communication (eMTC)\cite{Lin2017IoT, 3GPP2016IoT}. EC-GSM-IoT operates on legacy GSM bands and can support up to 240 kbps peak data rate over a 200 kHz channel. It applies advanced repetition and signal combining techniques to further extend the service coverage. % to up to 164 dB. % maximum coupling loss (MCL).} %with improved robustness.   %the new idle-extended-discontinuous-reception (I-eDRX) mode to further increase the battery life. It also supports improved security features over the GSM/EDGE systems. %It operates on 200 kHz per channel on traditional GSM bands.
NB-IoT is a new radio added to LTE. It focuses on low-end IoT applications. {For example, T-mobile recently announced plans to provide NB-IoT service at a rate of \$ 6 per year per device with up to 12 MB of data.} This service can achieve up to 250 kbps peak data rate over %supports narrowband operation with
180 kHz bandwidth on a GSM or LTE band, or on an LTE guard-band. eMTC is derived from LTE but with new power saving functions that can support up to 10 years of operation with a 5 Watt-hour battery. Due to its low transmit power, eMTC can coexist with high-speed LTE services. eMTC devices can support up to 1 Mbps data rate in both uplink and downlink over 1.08 MHz bandwidth. We summarize the main specifications of 3GPP IoT solutions in Table \ref{Tabel_IoTSolutions}. %fastest among all three IoT technologies.

\begin{comment}
The usage of licensed band resource must be centralized monitored and controlled following a time slotted synchronization protocol. This results in a tradeoff between battery life and the service latency. More specifically, to increase the battery life, LTE systems introduce the discontinuous reception (DRX) mechanism which allows each device to periodically check update of the system information broadcast by the network according to the DRX cycle. Each DRX cycle can range from 32 to 256 radio frames. Each device will decode the received system information update and request the channel connection if it identifies its any service request (e.g., receiving calls, messages, and connection requests). For example, a commonly used setting for DRX cycle is 128 frames which means that each device needs to wake up at every 1.28 seconds even in idle mode to checkin with the network. In other words, a significant amount of energy will be consumed by each device to receive and decode the system information even when there is no service request for most of the time.
\end{comment}

{To further improve the battery life of IoT devices, all IoT solutions adopt discontinuous reception (DRX) cycle, similar to LTE. In this setting, each device will periodically check the system information broadcast on the control channel according to the DRX cycle and only request a channel connection if it identifies a service request (e.g., receiving calls, messages, and connection requests). 
A typical LTE device can have up to 2.56 seconds of DRX cycle. 3GPP further extended the concept of DRX by introducing new extended discontinuous reception (eDRX) power saving modes for all three IoT solutions. In particular, two modes have been introduced for NB-IoT and eMTC: connected mode (C-eDRX) and idle mode (I-eDRX).} C-eDRX supports 5.12 seconds and 10.24 seconds of DRX cycles for eMTC and NB-IoT, respectively.  In I-eDRX, the DRX cycle can be further extended to 44 minutes and 3 hours for eMTC and NB-IoT, respectively. EC-GSM-IoT supports up to 52 minutes of DRX cycles. %These extended modes improve the battery life of IoT devices by sacrificing the service latency.

\subsection{Challenges for Massive IoT Deployment}

%Due to the high cost of licensed bands, to optimize the spectrum utilization, the
In spite of the strong push from industry and standardization organizations, many challenges remain to be addressed for massive deployment of IoT. % before real deployment and operation of IoT.

\subsubsection{Coexistence of Massive IoT and High-speed Cellular Services}
%The licensed band of each operator is resource limited and
%Resource-limited licensed band has already
%%Cellular networks have already overloaded with broadband mobile applications that require high data rate and low-latency services. It is therefore important to develop new coexisting mechanisms that can minimize the impact of the massive scale deployment of IoT devices on the cellular services. %Motivated by the fact that
%%
Motivated by the fact that IoT devices require low transmit powers and narrow bandwidth, % and therefore
most existing works focus on developing optimal power control, channel allocation, and  scheduling algorithms for IoT services to adapt to the dynamics of the coexisting cellular traffic. % to avoid causing intolerable interference to the existing cellular services.
However, IoT devices are usually low-cost with limited processing capacity to calculate and instantaneously adjust their transmit powers and channel usage. Some recent works suggest deploying edge/nano-computing servers at the edge of the network, e.g., BSs, to collect the necessary information and make decisions for near by IoT devices\cite{Zhang2017FogComputing}.
%One solution to address this issue is to deploy mobile edge computing, also called fog computing, to support IoT services. MEC will be a key enabler for the IoT implementations. In particular, the edge/nano computing servers deployed at the edge of the network, e.g., the BSs, can collect the necessary information and make decision for IoT devices.
These solutions make optimal resource allocation and instantaneous interference control possible for IoT devices. {However, deploying new infrastructure such as edge servers, enhanced/upgraded base stations, and new interfaces to support coordination and information exchange between BSs and edge servers requires extra investment from operators. For example, recent announcements from AT\&T and Verizon revealed that billions of dollars are required to upgrade their infrastructure for supporting IoT-based 5G networks. Such investment will eventually be  reflected in higher charges to end users. % on the prices/bills charged to the IoT service users operated on the 5G network infrastructure.}

\subsubsection{{Excessive Overhead and Inefficiency of Random Access Channel Procedure}}
{Another issue is that the random access channel (RACH) procedure currently used in LTE and GSM incurs high energy consumption and a significant amount of signaling overhead to establish connections between devices and network infrastructure. Directly extending this procedure to IoT systems is uneconomic and unrealistic.}
%
%for operating IoT in cellular systems is that  signalling overhead for each IoT device to develop connection with the network infrastructure is excessive and energy-consuming. 
In particular, it has been reported that in a typical cellular system, transmitting 100 bytes of payload from a mobile device requires  up to 59 bytes and 136 bytes of overhead on the uplink and downlink, respectively\cite{Shariatmadari2015MTC}.  %the resource allocation for licensed band follows a
%In addition, 
%This is because 
In addition, the RACH procedure %used in  %In particular, to maintain the required QoS, cellular systems follow
%most cellular standards, such as 
%LTE 
was originally designed to support only a limited number of mobile devices (around 100 mobile devices per cell). For example, if a device tries to establish a connection, it must %deploying a massive number of IoT devices (support up to 50,000 devices per cell in 3GPP standard) under this procedure introduces many new challenges. More specifically, in RA procedure, the BS periodically broadcasts timing and configuration information of the uplink and downlink carriers. Each device must periodically decode these information and
randomly choose a preamble signal sent to the BS over the physical random access channel (PRACH). %% This preamble signal will be decoded by the BS to evaluate the channel condition and allocate the appropriate resource for data communication. However,
%
%This introduces many new challenges when a large number of IoT devices is connected to each BS. %First, in RA procedure, each device is required to periodically check-in with the network. This results in a tradeoff between battery life and service latency. For example, a mobile phone needs to synchronize with the cellular network every 1.5 seconds even when it is out of use to maintain low latency.  The second challenge for RA procedure is to efficiently allocate resources to a large number of IoT devices. One of the main objective of 3GPP's IoT solutions is to   %All the above three technologies support low-power mode and are ready for
%support the deployment of IoT devices in a massive scale (up to 50,000 devices per BS). However,
%In particular, %in existing RA procedure, each IoT device needs to choose a preamble signal for the BS to allocate the requested resources.
%
%This protocol results in a huge increase for the possible network congestion, co-channel interference, and collisions during the RA especially when each BS has been associated with a large number of IoT devices. For example,
%More specifically,
In existing LTE systems, each device can only choose one preamble from a set of 64 pre-defined preamble signals. If two or more devices choose the same preamble, a conflict will happen which will result in retransmission and further delay in resource allocation. %To address these issues,  % and increased unreliability of data communication. % that cannot be tolerated for mission-critical IoT services. % preamble retransmission  the  Also allowing each IoT device to periodically decode and synchronize with the associated BS will also increase the energy consumption and reduce the battery life of IoT devices.

\subsubsection{Diverse QoS Requirements}
Another challenge related to the diverse requirements of IoT services is that existing IoT solutions treat data generated by different IoT services the same. In particular, for some massive-type IoT applications, such as long-term environmental monitoring and parcel tracking, a certain amount of data loss and data delivery latency can be tolerated. However, in mission-critical IoT applications, such as fire/gas alarm, health monitoring, and traffic safety, data delivery must be instant and highly reliable.
%It is known that 5G networks are expected to be future-proof and support diverse IoT services tailored for various systems and applications. %For example, some IoT application such as alarm signals can be delay-sensitive and require high reliability for data delivery.
%Existing access technologies treat each IoT device equally.
How to differentiate the service requirements for different applications and distribute appropriate resources to meet the needs of various IoT services is still an open problem.
%distribute the limited resource according to different service requirements on service latency, communication reliability, and battery life is still an open problem.  %This calls for a more flexible and

\subsubsection{Mobility Management and Traffic Dynamic Control}
Due to the mobility of UEs and IoT devices, as well as the time-varying traffic of different services, the resource demand and LTE/IoT coexisting topologies can be dynamic. Most existing solutions are focusing on optimizing the long-term performance based on a priori knowledge and/or prediction results. For example, an IoT device can predict the future change of its movement, change of data traffic as well as activities of other UEs in its proximity, so it can prepare for the future (e.g., scheduling/reserving a certain amount of bandwidth for future use if it predicts that these resources will soon be limited). However, always relying on each IoT device to predict its resource needs is impractical  due to the limited processing capability. Currently, there is no simple and economic solution that allows each IoT device to instantaneously adapt to the environmental dynamics without sacrificing the device's cost and battery life.    %at each IoT device is unrealistic. To address this issue, recent works suggested to deploy mobile edge computing infrastructure to optimize the service and application provision, placement, and scheduling for massive IoT devices\cite{Yu2018InfocomFogIoT}.

\section{Multi-operator Network Sharing for Massive IoT}
\label{Section_Liensed}

\subsection{Inter-operator Network Sharing Architecture}

%\begin{figure}
%\centering
%\includegraphics[width=5.3 in]{Figures/ActiveRANSharing.pdf}
%\caption{3GPP active radio access network (RAN) sharing architecture: (a) RAN-only sharing and (b) Gateway core network (GWCN). }
%\label{Figure_ActiveRANSharing}
%\end{figure}

%3GPP introduce two types of network sharing architectures: passive RAN sharing, also called infrastructure sharing, and active RAN sharing. %
%As mentioned earlier, IoT services will have diverse requirements and service demands which must be carefully considered when designing the resource allocation protocols of the pool.
%In this article, we introduce an inter-operator network slicing framework which allows two or more operators to share their network resources and jointly support a common set of different services with different QoS requirements. Our framework is built on the 3GPP's network sharing architecture. More specifically,
%
The concept of network sharing %has been introduced in 3GPP standards to allow operators to share the RAN resources, including network infrastructure and spectrum, with each other to reduce their system roll-out cost/delay.
%Network sharing
has been first introduced in 3GPP Release 10 to allow multiple operators to share their physical networks. Early development of network sharing mainly focused on infrastructure sharing, also referred to as {\em passive RAN sharing}\cite{3GPP2016NetworkShare, 3GPP2016NetworkShare2}. In this scenario,
%3GPP Release 14 introduces two network sharing architectures: passive and active RAN sharing \cite{3GPP2016NetworkShare}. In passive RAN sharing,
operators share site locations and supporting infrastructure such as power supply, shelters, and antenna masts. However, each operator still needs to install its own antennas and backhaul equipment for individual usage. %Recent development of network sharing in 3GPP Release 13 introduces the
3GPP Release 14 introduces the {\em active RAN sharing} architecture.
%In active RAN sharing, o
Operators can now share their spectrum resources as well as core network equipments (i.e., eNBs) based on a network sharing agreement, which can include mutual agreement on legal, finance, and joint operations. To ensure efficient and secure resource management, a master operator (MOP) is designated as the only entity that manages resource %, including both the core network and spectrum resources,
shared among the participating operators (POPs). %The resource usage of the spectrum pool will be managed by the master operator-network manager (MOP-NM).
The MOP can be a third-party manager designated by POPs. It can also be one of the POPs. % that has been agreed by others to act as the MOP to control and supervise the allocation of the shared resource. %Our framework extends from the active RAN sharing architecture to support the inter-operator spectrum sharing for massive IoT deployment.
In 3GPP's architecture, the MOP may charge POPs based on the requested data volume and the required QoS.

According to the entities shared by POPs, active RAN sharing architectures can be further divided into two categories:
\begin{itemize}
\item {\em RAN-only sharing}, also called multi-operator core network (MOCN). In here, a set of BSs sharing the same spectrum can be accessed by all POPs. Each POP, however, maintains its own core network elements, including the mobility management entity (MME) and serving and packet gateways (S/P-GW). Each POP can connect its core network elements to the shared RAN via the S1 interface.

\item {\em Gateway core network} (GWCN). In addition to sharing the same set of BSs. In GWCN, POPs can also share a common MME to further reduce costs.
\end{itemize}

%Multiple operators can contribute part of their licensed bands including both GSM and LTE bands to form a common spectrum pool.
%
%The licensed spectrum pooling has been introduced in 3GPP Release 14 as one of the main use case for 3GPP's network sharing architecture.

%One of the main advantage of supporting the IoT service in the multi-operator network sharing architecture is to reduce the cost of the IoT services. In particular, an IoT service provider only needs to purchase the licensed from one POP to access all the shared resources.
{To simplify the exposition, in the rest of this section, we assume that each POP corresponds to a cellular operator that  divides its network infrastructure and licensed spectrum into two parts: an {\it exclusive use} part that is reserved and exclusively used by itself, and a {\it shared} part that can be accessed by other operators. The shared parts of the infrastructure and spectrum of all the POPs are combined and managed by the MOP. Each IoT device or UE has already been assigned to a POP. The BSs of each POP need to calculate the channel reuse structure between the low-power NB-IoT devices and regular UEs so the cross-interference between both channel-sharing devices is below a tolerable threshold. In LTE, for example, the interference threshold for each UE is -72 dBm. If the exclusive use part of the spectrum is insufficient to support the traffic generated by the associated IoT and cellular services, the POP can temporally request a portion of shared spectrum from the MOP. If the spectrum requests of a POP are approved, the POP can assign any of its traffic (IoT or cellular) to the shared spectrum without consulting the MOP. If the spectrum requested by all  POPs exceeds the total amount of shared spectrum, MOP will partition  the shared spectrum and assign the divided spectrum to each POP according to a predetermined mutual agreement. }
Both active RAN sharing architectures can be extended to our framework to support spectrum sharing between IoT and high-speed cellular services. % as illustrated in Figure \ref{Figure_ActiveRANSharing}. 
In particular, RAN-only sharing allows each POP to adjust the traffic traversed through the shared network or its exclusive network resources according to the mobility of IoT devices (e.g., IoT services in wearable devices and vehicle networks). In this case, each POP needs to keep track of traffic dynamics and the required QoS requirements for both IoT services and its regular cellular services. The POP can then adjust the traffic sent through the shared infrastructure and its own exclusive infrastructure accordingly.
GWCN further reduces the cost for each POP by sharing a common MME among all POPs. It, however, cannot provide the same flexibility as RAN-only sharing for each POP because in this case mobility of devices is restricted to inter-RAN scenarios. In other words, each POP cannot adjust the traffic sent through the shared infrastructure and its own exclusive infrastructure by itself. In this case, POPs will need to predict the traffic from IoT services and cellular services, and reserve resources for each service accordingly.  %that can be requested by all the POPs. MON-NM will then schedule the proper portion of the resource (including antennas, transmission time, frequency bands) for each BS for each specific service. %To simplify our discussion, we consider the case that each IoT service provider has already subscribe the service of

\begin{figure}
\centering
\includegraphics[width=3.5 in]{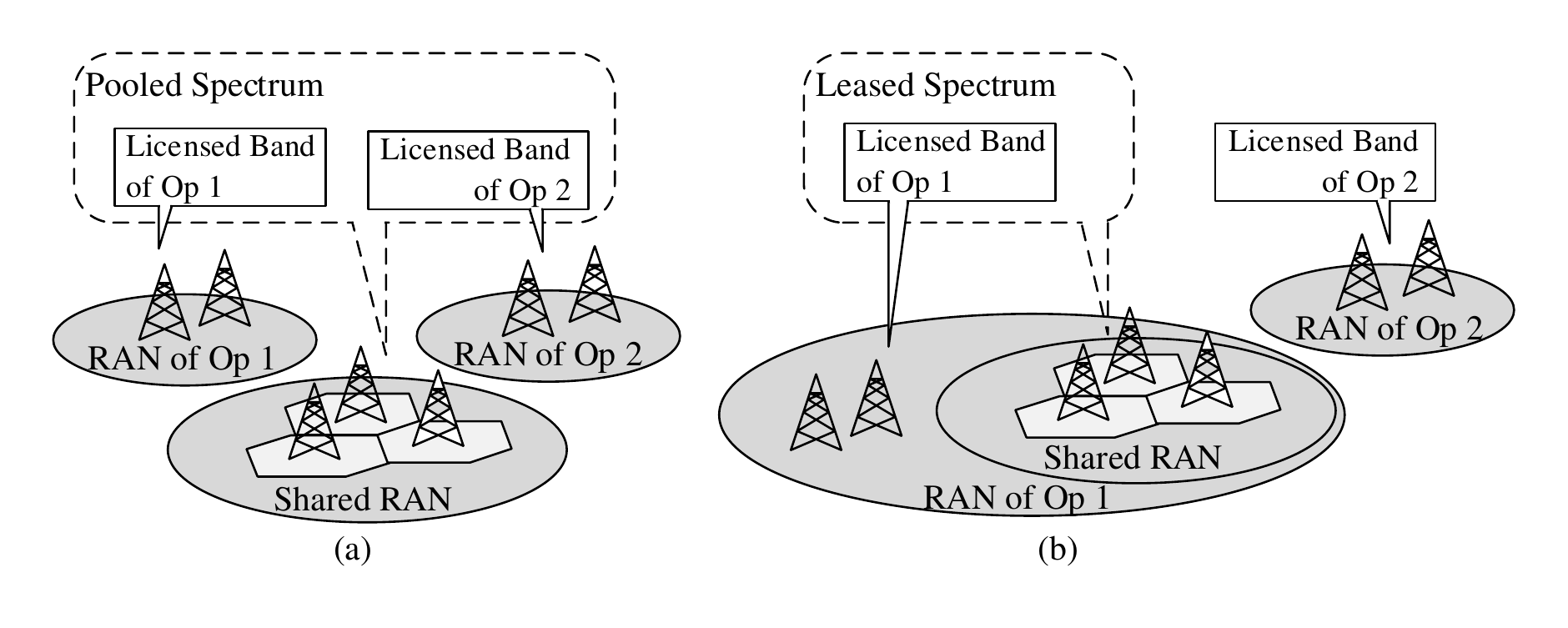}
\caption{Inter-operator network sharing: (a) spectrum pool, and (b) spectrum leasing.}
\label{Figure_SpectrumPoolandLease}
\end{figure}

%The spectrum used by the shared RAN cannot be changed when different POPs
In active RAN sharing, different POPs can access/rent different part of the shared infrastructure (e.g., a set of BSs that can be accessed by all POPs). However, the BSs in the shared RAN must operate on the same spectrum. Based on the spectrum used by the shared RAN, the multi-operator network sharing architecture can be further divided into the following two sub-categories, as illustrated in Figure \ref{Figure_SpectrumPoolandLease}:

\subsubsection{Spectrum Pooling} %% (also called {\em fully pooled radio resources} in 3GPP's specification\cite{3GPP2016NetworkShare})
POPs can merge their licensed (GSM and/or LTE) bands to form a common pool to be used by the shared RAN as shown in Figure \ref{Figure_SpectrumPoolandLease}(a)\cite{XY2013SpectrumPool}. Allowing the shared BSs to operate on the pooled spectrum can significantly reduce the complexity of spectrum management, i.e., it is uneconomic and too complex to allow each BS to switch its operational bands when it has been rented by different POPs. Spectrum pooling has been considered as one of the main use cases for the network sharing architecture in 3GPP's technical specification. In this architecture, each POP will need to coordinate with MOP's network management controller for channel assignment to avoid inter-cell interference between the BSs in the exclusive-use RAN and those in the shared RAN.
%This problem becomes more critical in RAN-only sharing because each POP will not share  %dynamically adjust the traffic and
%the mobility information for its IoT devices and cellular UEs with others.

\subsubsection{Spectrum Leasing} %% (also called inter-operator carrier aggregation\cite{XY2015IOCA})}
3GPP's architecture allows one of the POPs to serve as the MOP to manage and control the resource allocation of the shared RAN as shown in Figure \ref{Figure_SpectrumPoolandLease}(b). In this case, it is possible for one POP to lease a part of its BSs and the licensed band to be shared with other POPs. In spectrum leasing, to maintain the required QoS for the MOP, IoT and UEs associated with MOP can have the priority to access the shared spectrum. The POPs can only offload a limited traffic to the shared RAN if the resulting impact (e.g., throughput degradation) to the existing traffic of the MOP is below a tolerable level. %According to the duration of the spectrum leasing, the spectrum leasing can be divided into short-term leasing and long-term leasing.
If two or more POPs can lease their network infrastructure and licensed bands to each other at different time periods according to their traffic demands and resource availabilities, the spectrum leasing becomes equivalent to the {\em mutual renting} introduced in METIS' future spectrum system concept\cite{XY2015IOCA, Singh2015IOSS, XY2015D2D}

\subsection{Design Issues}

%The resource distribution for supported IoT services follows the following procedures. MOP-NM will first estimate the possible data traffic and transmission schedule of different IoT services. Since each IoT device has been
%and moderate and schedule the usage of the pool accordingly.

%\begin{figure}
%\centering
%\includegraphics[width=3.2 in]{Figures/SVvsNBS.pdf}
%\caption{Revenue division between two operators (op 1 and op 2) with {two fairness criteria: proportional fairness (PF) and Shapley value (SV)}.}
%\label{Figure_Fairness}
%\end{figure}

There are several important issues when deploying IoT services using our proposed multi-operator network sharing framework:

\subsubsection{Fair Revenue Division Among Operators for Spectrum Pooling}
%  Operators are profit-driven and therefore the first problem for LSP is to investigate the feasible conditions that could incentivize the spectrum pooling among operators. In \cite{XY2013SpectrumPool}, we have investigate the
%
%
In 3GPP's network sharing architecture, MOP can charge services (e.g., IoT services) using the shared resource according to the data usage and required QoS profiles. One intrinsic problem  is then how to divide the revenue obtained by MOP from serving IoT among all the resource-sharing POPs. This revenue division determines each POP's perception on the fairness of the sharing, and will in turn affect its willingness to share the licensed band with others. In other words, the revenue allocation must be fair in the sense that it needs to protect the interests of all the contributing operators and, more importantly, incentivize POPs to contribute their resources to the pool.
%Different operators have different investments in both licensed spectrum and infrastructure deployments.
In addition, to encourage operators with higher investment and more licensed spectrum resources to contribute, it must also take into consideration the contributions of different operators. In other words, operators that contribute more resources should have a larger share of the revenue from the pool. Various fairness criteria have been investigated for the spectrum pooling. In particular, in our previous work\cite{XY2013SpectrumPool}, we consider the scenarios that multiple operators form a spectrum pool and allow coexistence of
%to support both of
their cellular service and other low-power services (e.g., IoT services) in the same band as long as the resulting interference is less than a tolerable threshold. We prove that operators can use the price charged to the spectrum access of low-power services to control the admission of devices. We also investigate the fair revenue division between resource sharing operators. This framework can be directly extended to analyze coexistence of IoT (e.g., eMTC) and cellular services. In this case, the IoT traffic admitted to the spectrum pool will be controlled by the price of the MOP. %can also allow other systems to access the licensed pool by charging a certain amount of prices.
%The revenue obtained from supporting all the IoT services in the pool will be divided among operators.
%{In particular, we consider two popular fairness criteria for revenue division among operators: proportional fairness (PF) and Shapley value (SV) in Figure \ref{Figure_Fairness}. PF has been widely adopted in both wired and wireless network systems for resource allocation, scheduling and revenue division. The basic idea of PF is to distribute revenue by maximizing the utilization of the pool under the constraint of a minimum QoS guarantee for each POP. The SV is another important fairness criteria that distributes revenues among POPs according to their contributions. This feature is useful to incentivize the POPs to contribute more resource to the pool. In Figure \ref{Figure_Fairness}, we can observe that SV can always distribute the revenue according to the traffic admitted through the licensed bands of operators. PF, on the other hand, always equally divides the revenue.}

%\subsubsection{Inter-operator Spectrum Aggregation}
%In our previous work, we developed inter-operator spectrum aggregation to support  among operators by extending traditional carrier aggregation (CA) techniques developed in LTE Release 13 to a more general setting that supports
%inter-operator spectrum sharing and aggregation (IO-SA). Two IO-SA modes will be investigated: interleaved

\subsubsection{NOMA for Coexistence between Cellular UEs and Massive IoT} %Resource Sharing between IoT Services and Regular Cellular Services}
%As mentioned earlier, allowing massive IoT devices to access the same licensed bands that are currently used by regular cellular services introduces many challenges.
%
%It is commonly believed that the
As mentioned earlier, existing RA-based resource allocation approach cannot be applied to the IoT devices due to the physical limit of the licensed band and the inefficient design of the protocol. One possible solution is to apply non-orthogonal multiple access (NOMA). In particular, NOMA improves the  utilization of cellular spectrum %by allowing IoT and cellular services coexist in the same time and/or frequency 
by exploiting power and code domain reuse. It provides the operators with more flexibility to increase the number of channel sharing devices, e.g., each BS can carefully choose different numbers of low-power IoT devices and high-power UEs at different locations to share the same channel.
Furthermore, NOMA does not require IoT devices to perform RACH procedure for data transmission. In particular, in NOMA, the random access and data communication can be combined\cite{Ding2017NOMASurvey}. For example, each IoT device can randomly pick up a narrowband and start data transmission without waiting for the channel assignment from the BS. The BS can then perform successive interference cancellation to decode the message of each IoT device received in each frequency band. The authors in \cite{Shirvanimoghaddam2017NOMAIoT} suggested to apply rateless Raptor codes to generate as many coded symbols as required by each BS, so each BS can differentiate the message sent by different IoT devices. It has been observed that the more difference in channel gains between IoT/UE and the BS, the higher performance improvement can be achieved by the NOMA. %, compared to the traditional orthogonal multiple access approaches such as FDMA, TDMA, and CDMA.   % and can  Due to the low transmit power and narrow bandwidth required for each IoT transmission, it is possible to deploy adaptive control approaches such as power control, adaptive channel code rate to optimize the coexistence of IoT and cellular services. More specifically, the BS can allocate low-power IoT transmitters and high-power cellular UEs to share the same channel at the same time as long as

%Allocating each IoT device with a frequency band for exclusive use as the existing cellular services  will not be able to  Recent observation suggests that the existing orthogonal multiple access approaches such as FDMA, TDMA, and CDMA, will not be able to support massive IoT devices due to the resource limit of the licensed bands. NOMA has been considered as one of the promising solutions to allow coexistence of low-speed IoT services and fast-speed regular cellular services. Several solutions have been introduced to implement NOMA for massive IoT services.

%MOP needs to also collect information from

%NOMA can be better supported by GWCN because allowing MOP to manage the MME can ...

\subsubsection{Network Slicing for Diverse IoT services}
Network slicing is a concept recently introduced by 3GPP to further improve the flexibility and scalability of 5G. The main idea is to create logical partitions of a common resource (e.g., spectrum, antenna, and network infrastructure), known as the slices, to be orchestrated and customized according to different service requirements. Network slicing has the potential to significantly improve spectrum efficiency and enable more flexible and novel services that cannot otherwise be supported by the existing network architecture. In our previous work, we have proposed an inter-operator network slicing framework to support different services with different requirements on a commonly shared resource pool formed by multiple operators\cite{XY2018IONetworkSlicingJSAC}. In this framework, a software-defined mobile network controller will be deployed in the MOP's network infrastructure that can isolate and reserve a certain amount of resource for each type of IoT services (e.g., wearable IoT devices, machine-type IoT, and smart infrastructure).  The controller will predict the possible future traffic of all the supported IoT services and can adjust the portion of the resource reserved for each service.

%different IoT services according to their requirements. The revenue obtained from the IoT service provider will be divided among operators. To allocate appropriate resource to each IoT device, the MOP will divide the spectrum pool into smaller component carriers (CCs) in both regular LTE/GSM carriers and guard-bands. MOP-NM can assign a single CC or multiple CCs to each IoT device at a time.

%The IoT devices can also access the CCs that are currently used by other LTE/GSM UEs. In this case, the MOP-NM will monitor the cross-interference between IoT devices and UEs and carefully assign the transmit power for both UEs and IoT devices.

\begin{comment}
billing share/division
interference manage
scheduling

\subsection{Licensed Spectrum Renting}
In particular, in 3GPP's architecture, the MOP can also be one of the POP.

\subsubsection{Dynamic Inter-operator Spectrum Aggregation}

\subsubsection{Dynamic Coordination and Scheduling}

\subsubsection{Inter-operator Interference Control}

\subsection{Unlicensed Spectrum Access Sharing}

\end{comment}

\section{Performance Evaluation} % Multi-operator Spectrum Sharing}

\begin{figure}
%\begin{minipage}[t]{0.65 \linewidth}
\centering
\includegraphics[width=6.2 in]{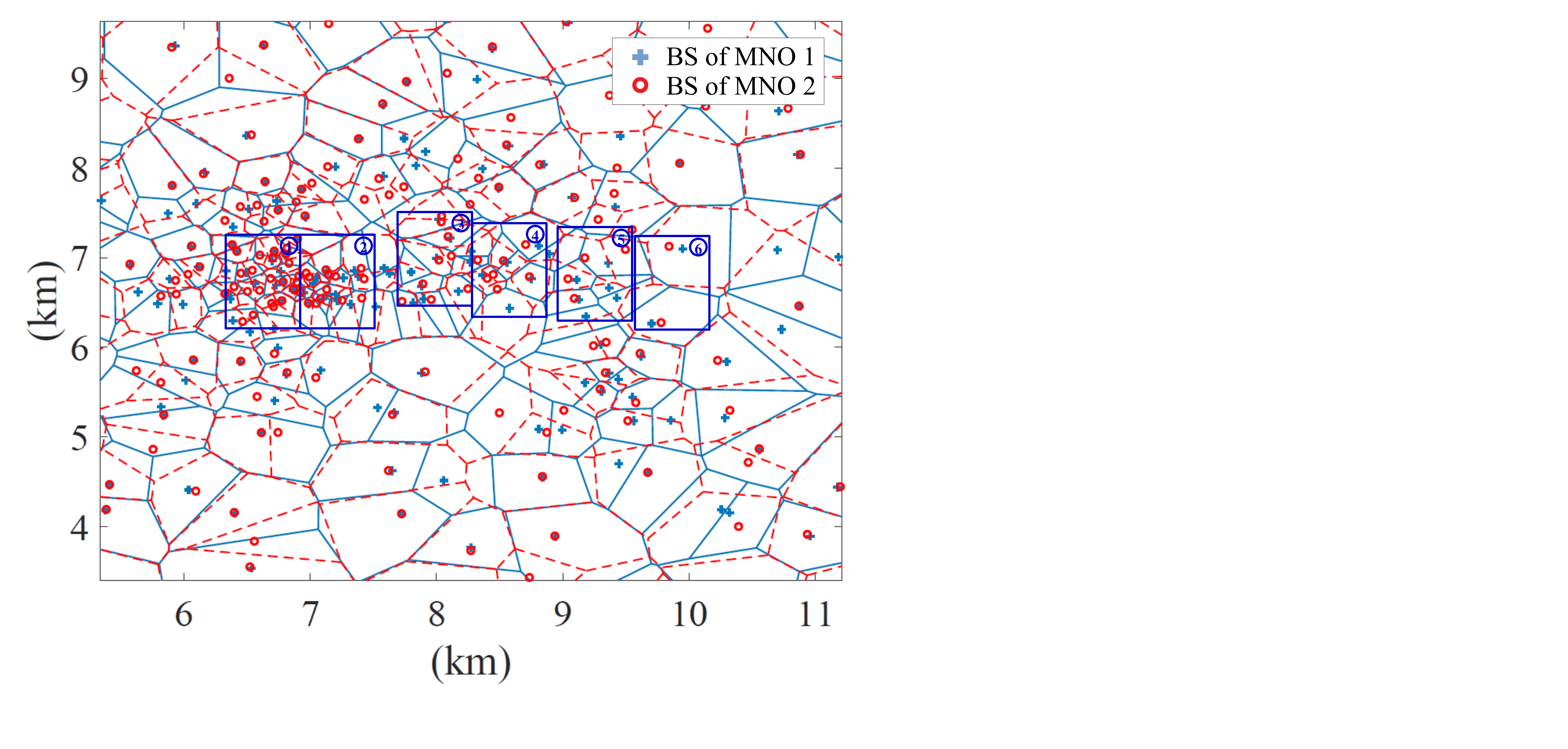}
\vspace{-0.15in}
%\caption{(a) Locations of BSs deployed by two major cellular operators in the city of Dublin (b) average cell sizes in the considered areas. }%\vspace{-0.3in}
%\end{figure}
%\end{minipage}
%\begin{minipage}[t]{0.3\linewidth}
%%\begin{table}[tbp]
%\vspace{-1.8in}
%\centering
%\label{tb:sim_parameters}
%\vspace{0.1in}
%%\scriptsize
%%\small
%\tiny
%%\resizebox{\columnwidth}{!}{
%\begin{tabular}{|l|l|}
%\hline
%Parameter & Value \\
%\hline \hline
%BS Transmit Power & $23$ dBms \\
%\hline
%IoT Transmit Power & $20$ dBms \\
%\hline
%UE noise floor& $-100$ dBm\\
%\hline
%IoT noise floor& $-90$ dBm\\
%\hline
%Path Loss Model& $43.3\log(d) + 11.5 + 20\log(f_c)$\\
%\hline
%Shadow fading& Log-normal ($\mu = 1$ dB, $\sigma^2 = 4$ dB)\\
%\hline
%Interference threshold& $-62$ dBm \\ 
%\hline 
%\multicolumn{2}{c}{} \\
%\multicolumn{2}{c}{} \\
%%\newline(c)%
%%\newline
%\end{tabular}
%%\newline \hfill
%{(c)}
%%\end{table}
%\end{minipage}
%\caption{(a) Locations of BSs deployed by two major cellular operators in the city of Dublin, (b) average cell sizes in the considered areas, and (c) simulation parameters. }%\vspace{-0.3in}
\caption{Locations of BSs deployed by two major cellular operators in the city of Dublin. }%\vspace{-0.3in}
\label{Figure_BSlocations}
\end{figure}

\begin{figure}
\centering
\includegraphics[width=3.5 in]{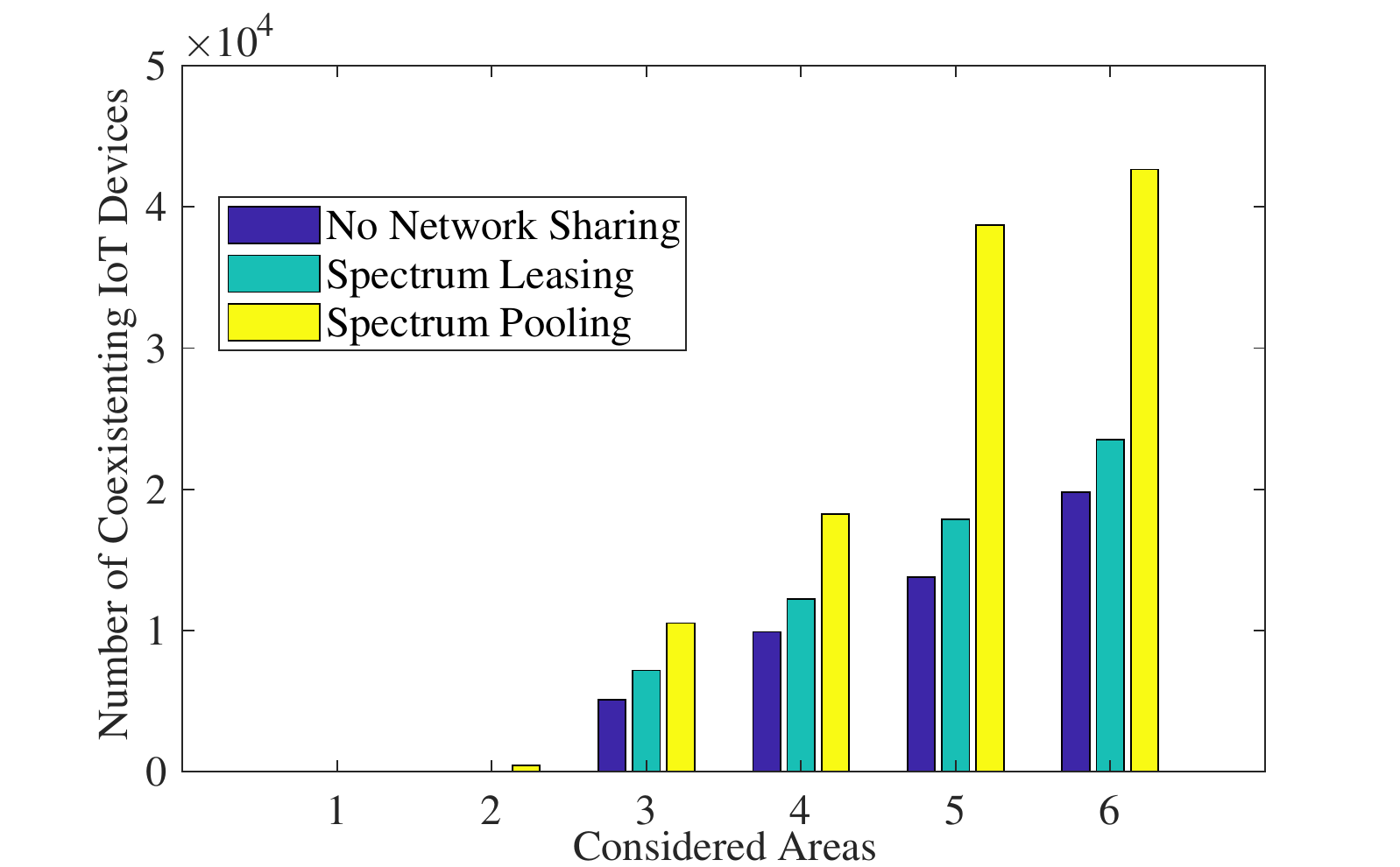}
\vspace{-0.15in}
\caption{Maximum number of IoT (eMTC) devices that can be coexisted with cellular UEs in different considered areas. }%\vspace{-0.3in}
\label{Figure_IoTNumberVSAreas}
\end{figure}

\begin{figure}
\centering
\includegraphics[width=3.5 in]{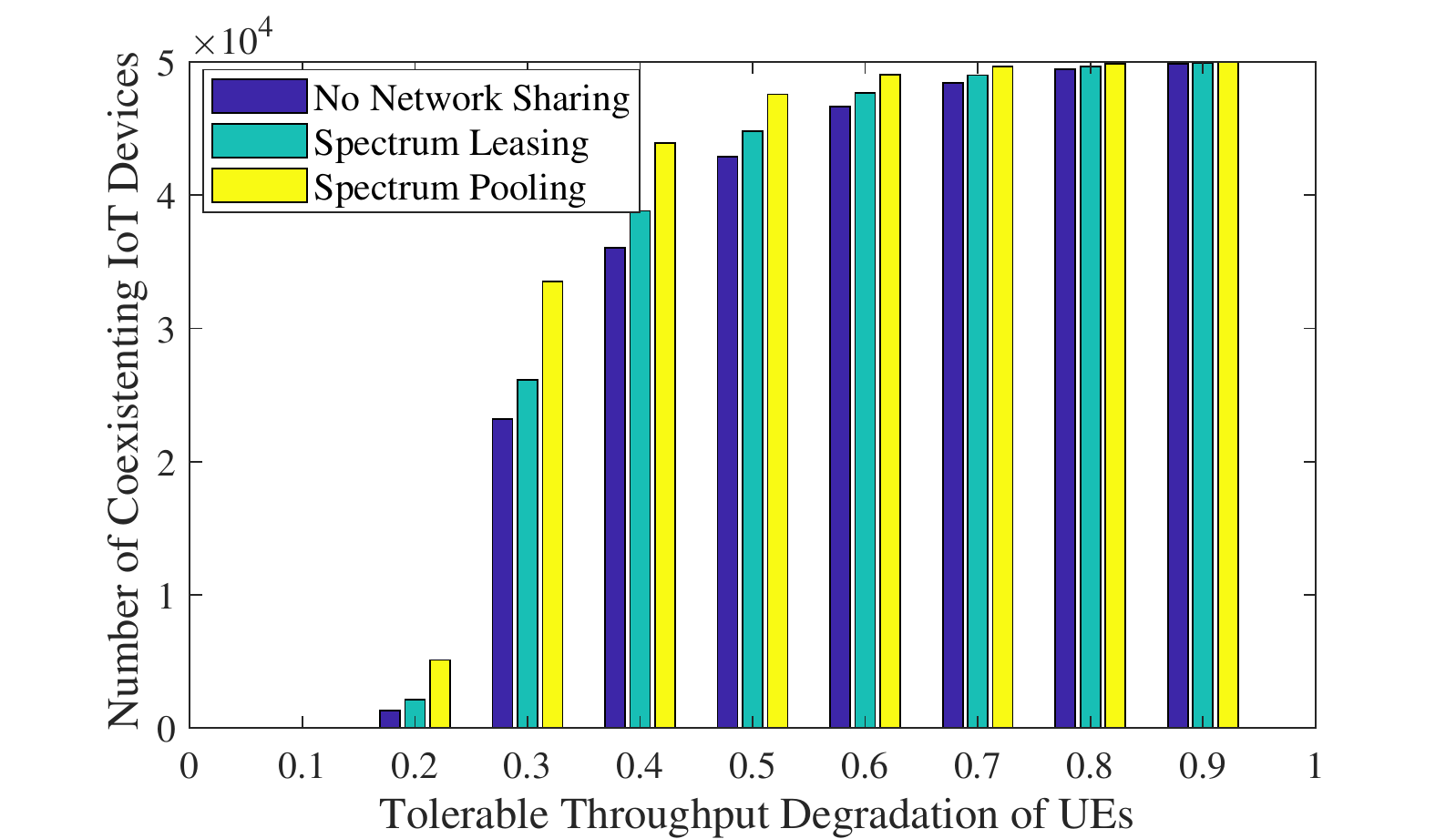}
\vspace{-0.15in}
\caption{Maximum number of IoT (eMTC) devices that can be coexisted with cellular UEs under various tolerable throughput degradation of UEs. }%\vspace{-0.3in}
\label{Figure_IoTNumberVSRateDeg}
\end{figure}

%simulate a multi-operator network sharing architecture using real BS location information from two major telecom operators in Ireland. We consider the saturated traffic For both operators, we consider saturated traffic and derive the maximum performance improvement that can be achieved by our architecture. 

To evaluate the performance improvement that can be achieved by our framework, {we simulate a multi-operator network sharing architecture using over 200 real BS locations in the city of Dublin deployed by two major telecom operators in Ireland. The actual distribution and the deployment densities of BSs are shown in Figure \ref{Figure_BSlocations}. We consider saturated traffic for both UEs and IoT devices and evaluate the possible coexistence of IoT (e.g., eMTC) devices and cellular UEs for uplink data communication in the same LTE band. Our results can be regarded as the maximum performance improvement that can be achieved by multi-operator network sharing architecture.   %We consider saturated traffic and assume each IoT device and cellular UE will always connected to the closest BS. We set the
The transmit powers of each IoT device and cellular UE are set to 20dBm and 25dBm, respectively. We assume 20 UEs and 50,000 IoT devices are uniformly randomly located in each cell. Each UE occupies a 5 MHz bandwidth. Each IoT device is randomly allocated with a 1 MHz bandwidth channel and can only send data with 20 dBm of transmit power. %We assume each  %We assume the UEs and IoT are with randomly allocated LTE bands.
IoT devices can only be supported when the interference to the UEs is lower than the LTE tolerable interference threshold (-62dBm).} %s interference to the UEs causes a tolerable depredation in the throughput. We compare the maximum number of IoT devices that can coexist with the UEs with and without network sharing. %In particular, we assume the saturated traffic for both UEs and IoT devices and each UE can only tolerate a limited throughput degradation caused by the co-channel interference from the IoT transmission.

In Figure \ref{Figure_IoTNumberVSAreas}, we carefully select 6 areas from the city center to suburban areas (representing different sizes and deployment density of cells) and compare the maximum number of IoT devices that can simultaneously transmit data with the UEs in the same LTE bands when each UE can tolerate 10\% of throughput degradation. We observe that when the size of the cell is small, the number of IoT devices that can share the same spectrum as the UEs is limited due to the high cross-interference between IoT devices and cellular UEs. However, as the size of the cell increases, the total number of coexisting IoT devices can increase significantly. In addition, allowing both operators to share their spectrum via pooling can almost double the total number of IoT devices when the deployment density of BSs is low. {This result complements the existing efforts of 3GPP on promoting the network sharing for 5G networks and could have the potential to influence the future practical implementation of the network sharing architecture between major operators.} % supporting a wide variety of services including both high-speed broadband services as well as low-throughput IoT services. 

In Figure \ref{Figure_IoTNumberVSRateDeg}, we compare the maximum number of IoT devices that can share the same channel with UEs when throughput degradations that can be tolerated by the each UE are different. We observe that the number of IoT devices increases when the UEs can tolerate a higher degradation for their throughput. In addition, network sharing provides more improvement in coexisting IoT traffic when the UEs can only tolerate a small throughput degradation, i.e., network sharing can almost double the maximum number of coexisting IoT devices when each UE can tolerate 20\% throughput degradation. However, when the tolerable throughput degradation of UEs increases to 90\%, the total number of coexisting IoT devices approaches the maximum values even without network sharing. %This verifies our previous observation that
In other words, network sharing can provide more performance improvement when the UEs require a stringent QoS guarantee with a limited interference tolerance.

\section{Conclusion}
\label{Section_Conclusion}
%This article has presented an overview of DET and its possible implementations into the paradigm of wireless powered communication systems.

In this article, we reviewed the current IoT solutions introduced by 3GPP. %We then  the 3GPP network sharing architecture to support massive IoT. %We first provide an overview of the recent 3GPP standardizations on IoT solutions and discuss the challenges for massive deployment of the IoT. %We then introduce the
%We describe the main difference between the existing intra-operator spectrum sharing and the LSS, and then compare these technologies with the ULL. We highlight the spectrum sharing strategies and control policies proposed for both LSS and ULL as well as the recent standardization progress and other implementation issues.
We then introduced a multi-operator network sharing framework based on 3GPP's network sharing architecture to support coexistence of massive IoT and regular cellular services offered by multiple operators.
%Our proposed framework is based on the active RAN sharing architecture recently introduced in 3GPP Release 13.
Various design issues were discussed. Finally, we simulated a multi-operator network sharing scenario using real BS location data provided by two major operators in the city of Dublin. Our numerical results show that our proposed framework can almost double the transport capacity of coexisting IoT traffic under certain scenarios.
%We also highlight some of the future research directions at the end of this article.
%To the best of our knowledge, this is the first work that discuss the deployment of IoT solutions on multi-operator network sharing architecture.

\section*{Acknowledgment}
The authors would like to thank Professor Luiz A. DaSilva and Dr. Jacek Kibilda at CONNECT, Trinity College Dublin for providing the BS location data in the city of Dublin. M. Krunz was supported in part by NSF (grants \# IIP-1822071, CNS-1563655, CNS-1731164) and by the BWAC center. T. Shu is supported in part by NSF (grants \# CNS-1837034, CNS-1745254, CNS-1659965, CNS-1659962, and CNS-1460897). Any opinions, findings, conclusions, or recommendations expressed in this paper are those of the author(s) and do not necessarily reflect the views of NSF. 

% if have a single appendix:
%\appendix[Proof of the Zonklar Equations]
% or
%\appendix  % for no appendix heading
% do not use \section anymore after \appendix, only \section*
% is possibly needed

% use appendices with more than one appendix
% then use \section to start each appendix
% you must declare a \section before using any
% \subsection or using \label (\appendices by itself
% starts a section numbered zero.)
%

%\appendices
%\section{Proof of the First Zonklar Equation}
%Appendix one text goes here.
%
%% you can choose not to have a title for an appendix
%% if you want by leaving the argument blank
%\section{}
%Appendix two text goes here.

% use section* for acknowledgement
%\section*{Acknowledgment}
%The authors would like to thank...

% Can use something like this to put references on a page
% by themselves when using endfloat and the captionsoff option.
%\ifCLASSOPTIONcaptionsoff
%  \newpage
%\fi

% biography section
%\centering
\bibliography{reference}
\bibliographystyle{IEEEtran}

\begin{IEEEbiographynophoto}{Yong Xiao}(S'09-M'13-SM'15) is a professor in the School of Electronic Information and Communications at the Huazhong University of Science and Technology (HUST), Wuhan, China. He received his B.S. degree in electrical engineering from China University of Geosciences, Wuhan, China in 2002, M.Sc. degree in telecommunication from Hong Kong University of Science and Technology in 2006, and his Ph. D degree in electrical and electronic engineering from Nanyang Technological University, Singapore in 2012. His research interests include machine learning, game theory, distributed optimization, and their applications in cloud/fog/mobile edge computing, green communication systems, wireless networks, and Internet-of-Things (IoT). 
\end{IEEEbiographynophoto}

\begin{IEEEbiographynophoto}{Marwan Krunz}(S'93-M'95-SM'04-F'10) is the Kenneth VonBehren Endowed Professor in ECE at the University of Arizona. He is the center director for BWAC, a multi-university NSF/industry center that focuses on next-generation wireless technologies and applications. He previosuly served as the site director for  the Connection One center. Dr. Krunz’s research emphasis is on resource management, network protocols, and security for wireless systems. He has published more than 280 journal articles and peer-reviewed conference papers, and is a co-inventor on several patents. He is an IEEE Fellow, an Arizona Engineering Faculty Fellow, and an IEEE Communications Society Distinguished Lecturer (2013-2015). He received the NSF CAREER award. Currently, he serves as the Editor-in-Chief for the IEEE Transactions on Mobile Computing. He previously served as editor for numerous IEEE journals. He was TPC chair INFOCOM’04, SECON’05, WoWMoM’06, and Hot Interconnects 9. He was the general vice-chair for WiOpt 2016 and general co-chair for WiSec’12.
\end{IEEEbiographynophoto}

\begin{IEEEbiographynophoto}{Tao Shu} received the B.S. and M.S. degrees in electronic engineering from the South China University of Technology, Guangzhou, China, in 1996 and 1999, respectively, the Ph.D. degree in communication and information systems from Tsinghua University, Beijing, China, in 2003, and the Ph.D. degree in electrical and computer engineering from The University of Arizona, Tucson, AZ, USA, in 2010. He is currently an Assistant Professor in the Department of Computer Science and Software Engineering at Auburn University, Auburn, AL. His research aims at addressing the security, privacy, and performance issues in wireless networking systems, with strong emphasis on system architecture, protocol design, and performance modeling and optimization. 
\end{IEEEbiographynophoto}

%\newpage

% that's all folks
\end{document}